\documentclass[lettersize,journal]{IEEEtran}
\usepackage{amsmath}
\usepackage{amsfonts}
\usepackage{amssymb}
\usepackage[ruled,vlined]{algorithm2e}
\usepackage{caption}
\usepackage{mathtools}  
\usepackage{amsthm}
\usepackage[]{algorithm2e}
\usepackage{booktabs}
\usepackage{subfigure}
\usepackage{tabulary}
\usepackage{footnote}
\usepackage{caption}
\usepackage[export]{adjustbox}
\usepackage{booktabs}
\usepackage[justification=centering]{caption}
\usepackage{comment}

\usepackage{cases}
\usepackage{graphicx}
\usepackage{cite}
\usepackage{multicol}
\usepackage{xcolor}
\usepackage{graphicx}
\graphicspath{{./}{figures/}}
\newcommand\numeq[1]%
  {\stackrel{\scriptscriptstyle(\mkern-1.5mu#1\mkern-1.5mu)}{=}}
\usepackage{stfloats}
\usepackage{cuted}
\usepackage{lipsum}

\usepackage{subfigure}
\usepackage{multirow}
\usepackage{mathtools}
\DeclarePairedDelimiter\ceil{\lceil}{\rceil}
\DeclarePairedDelimiter\floor{\lfloor}{\rfloor}

\title{A Light-Weight Communication-Efficient Data Sharing Approach in 5G NR V2X} 

\author{\text{Ran Wei}, \text{Lyutianyang Zhang} \thanks{
Ran Wei and Lyutianyang Zhang are with Department of Electrical \& Computer Engineering, University of Washington, Seattle, WA, USA (e-mail:\{rawe1722, lyutiz\}@uw.edu).}}%

\begin{document}
\maketitle
\begin{abstract}

Timeliness of information is critical for Basic Safety Messages (BSMs) in Vehicle-to-Everything (V2X) communication to enable highly reliable autonomous driving. However, the current semi-persistent scheduling (SPS) algorithms in the 5th generation  New Radio (5G NR) standard can still generate collisions probability, close to 20\% with 100 vehicles per kilometer, such that they cannot meet this requirement for BSMs.This paper proposes a Ledger concept, which can communicate collision information to every vehicle in the network within one Resource Reservation Interval (RRI) through the broadcasting of Ledger information. The colliding vehicle is aware that it has collided with other vehicles and will reselect for the next SPS period. Besides that, other protocols follow the SPS. Although it sacrifices 14.29\% of resources by including Ledger, it can eventually reduce the collision probability. In this paper, a Monte Carlo simulator is used to verify and analyze the performance of the Ledger system. Numerical results show that abide by the SPS protocol, the Ledger system can make the collision probability converge to zero after amount of RRIs. 

\end{abstract}
\begin{IEEEkeywords}
5G, Vehicle to Everything, Block Chain, Distributed Ledger, Resource Allocation
\end{IEEEkeywords}

\section{Introduction}
Safety-related applications, such as the Basic Safety Message (BSM), are part of the regular vehicle status communication. The purpose of BSM is to assess potential road hazards by announcing the presence of a vehicle to other vehicles around it. In order for the vehicle status to be fresh enough, BSM scheduling is crucial. 3GPP Release 16 introduced NR-V2X \cite{release16}. Mode 2 in Release 16 is a more suitable communication mechanism for use in a V2V environment that uses sensor-based Semi-Persistent Scheduling (SPS)\cite{9461188}. It is a strategy for a decentralized direct communication protocol for broadcast. However, this algorithm may not meet the requirements of timeliness and reliability. Liu \emph{et al.} in \cite{9771765} studied that collision probability close to 20\% with 100 ms RRI and 100 vehicles per kilometers. The most worse case is that the BMS packages are delayed by 1.5 seconds \footnote{If RRI = 100 ms, Re-selection Count (RC) = 15, the collision happened in one SPS = RC * RRI = 15 * 100 = 1500 ms = 1.5 s.}, which is dangerous enough in case of an actual emergency.

Before the blockchain technology introduced in 1991, the distributed ledger technology (DLT) conceptually arose in 1982. Distributed ledger technology  has established itself as an umbrella term to designate multi-party systems that operate in an environment with no central operator or authority, despite parties who may be unreliable or malicious (‘adversarial environment’) \cite{rauchs2018distributed}.

Cellular-V2X (C-V2X) was introduced by 3GPP in Release 14 and was competing with Dedicated Short Range Communications (DSRC) \cite{molina2017lte,cao2020performance}. It is currently move to NR-V2X. The authors in paper \cite{feng2019predictive} developed a predictive SPS scheme that utilizes transmission history to effectively reduce the uplink latency of LTE systems and NR systems. Based on an Age-of-Information (AoI) perspective, the paper \cite{9771765} examines the parameters of the SPS used in NR-V2X Mode 2 for BSM scheduling. The authors in \cite{9527765,cao2022resource} discussed to apply the deep reinforcement learning algorithm under the platooning scenario to increase the resource allocation efficiency. \cite{dayal2021adaptive} introduces SPS++ to accommodate an adaptive RRI. This system allows each vehicle to dynamically adjust RRI to decreasing the collision probability.

To mitigate these drawbacks, this paper utilized DLT to facilitate data sharing (resource occupation) in a distributed manner, especially collision information. Distributed Ledgers serve as a lightweight database. Each node should periodically share its information (such as occupied sub-channel ID, transmission subframe timestamp, collision sub-channel ID, its subframe timestamp, and so on) with the network. In a few microseconds, every node in the network receives all newly added data if no collision happened, and collision information if there is a collision \cite{distributed2017}. Therefore, applying such technique will allow each vehicle to establish a common, transparent, irreversible, distributed and lightweight Ledger \cite{burkhardt2018distributed,cao2021blockchain}. Within one SPS period, each vehicle could know if another vehicle collided with it or not, based on the shared Ledger. 

Thus, this paper aims to introduce a novel mechanism for resource allocation in broadcast V2V networks. This mechanism will be based on the SPS mechanism and will also be compatible with the 5G NR V2X protocol. In a sense, full-duplex mode is implemented. Simulation results showed a substantial reduction in collision and continuous collision.

\begin{table*}[t]
  \centering
  \caption{Numerologies in NR V2X SL}
  \label{tb:SCS}
  \begin{tabular}{|c|c|c|c|c|c|c|c|}
    \hline
    \text{$\mu$} & \text{SCS [kHz]} & \text{$N_{slot}^{frame,\mu}$} & \text{$N_{slot}^{subframe, \mu}$} & \text{$\tau_{slot, \mu}$ [ms]} & \text{$N_{symbol}^{slot}$} & \text{$N_{symbol}^{subframe, \mu}$} & \text{Maximum Carrier Bandwidth [MHz]} \\ \hline

    0 & 15 & 10 & 1 & 1 & 14 & 14 & 50 \\ \hline
    1 & 30 & 20 & 2 & 0.5 & 14 & 28 & 100 \\ \hline
    2 & 60 & 40 & 4 & 0.25 & 14 & 56 & 200 \\ \hline
    3 & 120 & 80 & 8 & 0.125 & 14 & 112 & 400 \\ \hline
 \end{tabular}
\end{table*}

\section{System Model}

 \begin{figure}[t]
    \centering
    \includegraphics[scale=0.11]{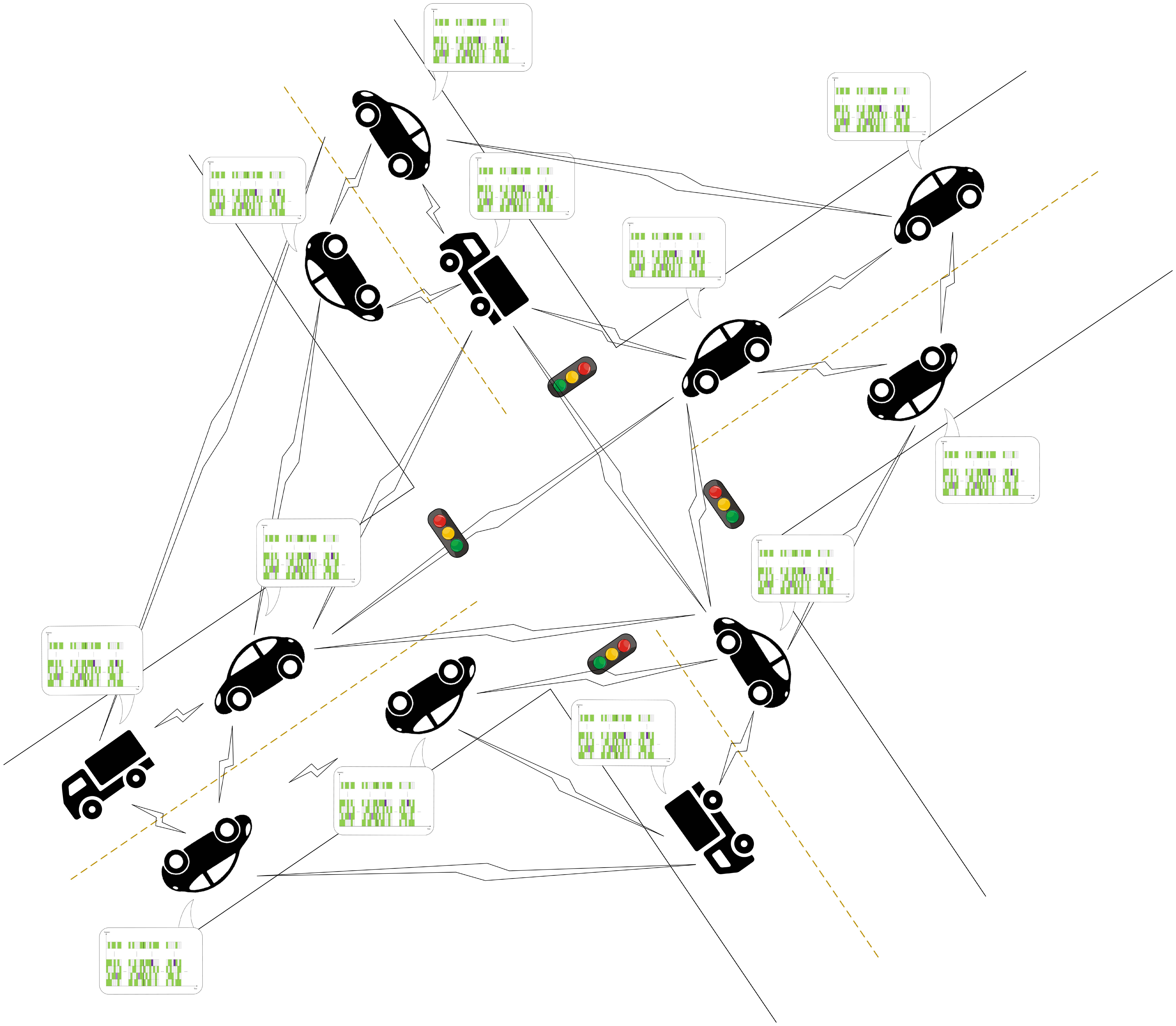}
    \caption{Application Scenario.}
    \label{fig:scenario}
\end{figure}

\begin{algorithm}
\caption{Steps of Establishing the Ledger}
\label{Ledger}
\KwData{Sensing list $L_A$, received package}
\KwResult{Establish Ledger package for each subframe}
initialization buffers;\\

\While{join the current V2V network}{
    sensing current subframe channel;\\
    \eIf{not the RRI subframe}{
        receive package from other vehicles;\\
        decode the package and add to local Ledger;\\
        
        \If{any sub-channel energy high but non decoded package match it}{
        record collision sub-channel ID;\\
        record corresponding subframe timestamp;
        }
    }{
        transmit the package;
    }
}
\end{algorithm}

Generally, vehicles transmit and receive packages in half-duplex (HD) mode. Due to Half-Duplex, at the transmission subframe, the vehicle cannot hear other vehicles. It cannot build up the Ledger for that subframe. With this Ledger design, full-duplex mode is realized with some delay. First, the HD condition means the vehicle cannot receive any packages from other vehicles in the same slot. With Ledger sharing, the transmission vehicle could fill in the missing package information in its transmission slot. It ensures that every vehicle in the network has the same Ledger. In addition, the vehicle in the network always senses the entire channel due to the SPS protocol. Therefore, every vehicle could compare with the sensing list after receiving and decoding the packages in each slot. The collision sub-channel ID could be detected and included into the transmission package for sharing. It ensures that collision information can be transmitted as soon as possible. 

Based on the scenario shown in Fig. \ref{fig:scenario}, each vehicle in the network has a fixed algorithmic program to maintain the operation of the entire new system model. This redesigned system model then contains the Ledger concept proposed in this paper. All communications are compatible with the 5G NR V2X SPS protocol.

The algorithm requires two types of data, the sensing list $L_A$ and the received packages from other vehicles in the network. The final goal of the algorithm is to build the complete Ledger.

There are some buffers that used to record different information. Therefore, each buffer must be initialized before use them. For instance, the newest sensing list $L_A$ should be loaded into buffer. Once the vehicle joins the existing V2V network, it will always follow the SPS protocol and will continue to sense the current subframe. The current subframe list $L_A$ is the result. The $L_A$ list is a record of the occupancy of each sub-channel of the entire channel at the time of the current subframe. This occupancy is represented by the energy of the sub-channel. The RSRP and RSSI thresholds based on SPS protocol are used to classify which sub-channels are busy (above the threshold) and which sub-channels are idle (below the threshold). In the case of a collision, the sub-channel with the collision will also carry energy, so it will also be listed as busy in the $L_A$ list.

The vehicle is selected at the beginning of the SPS period to broadcast the package in a fixed sub-frame of each RRI according to the protocol. Whenever it is in the subframe of a RRI where the package is to be delivered, the vehicle broadcasts the corresponding stored package. In the case of a subframe not needed for sending, the vehicle must collect and analyze the information to assemble a package for the next sending subframe.

\begin{figure*}
  \centering \includegraphics[scale=0.23]{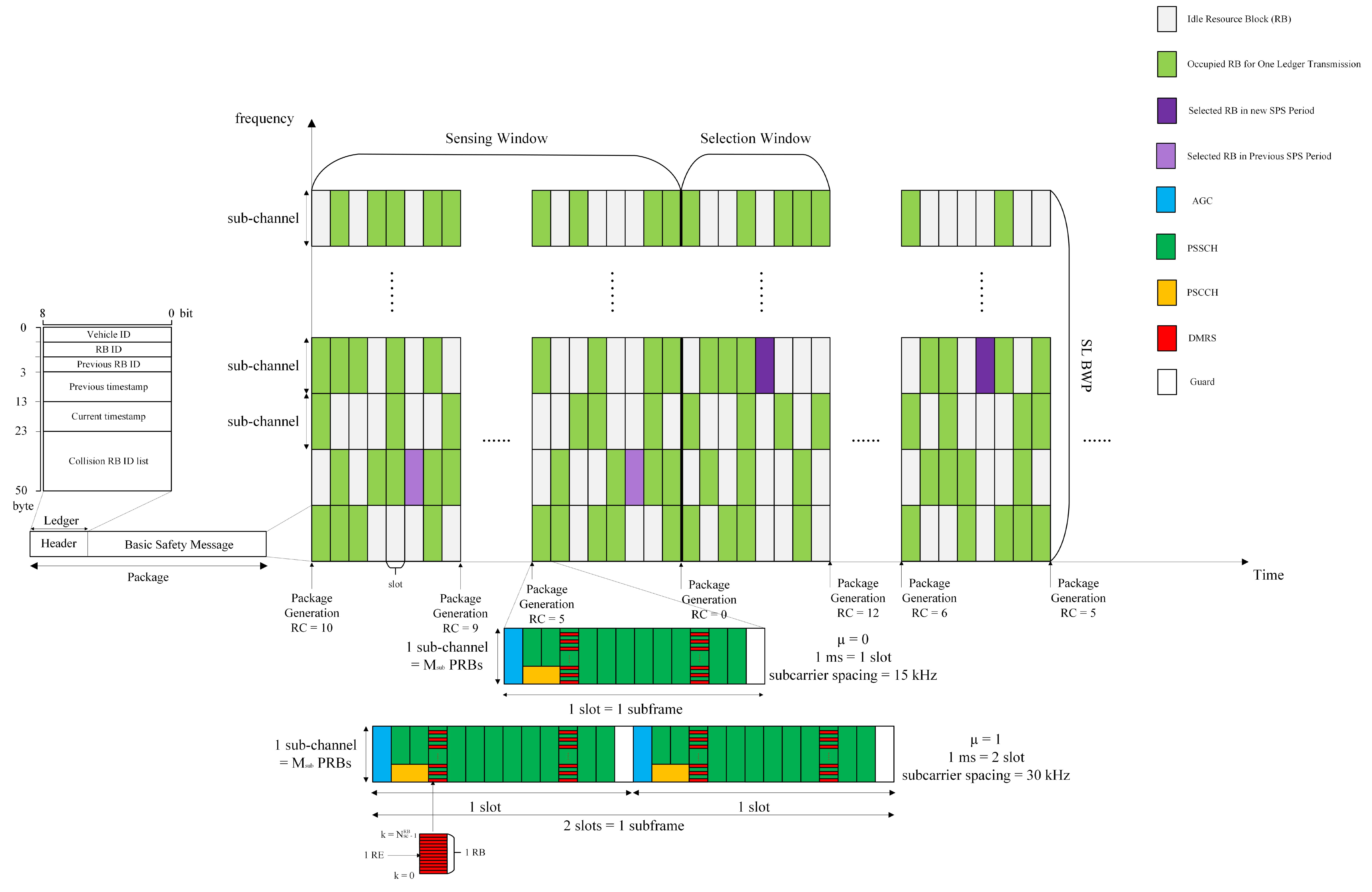}
  \caption{Ledgers and physical layer.}
  \label{fig:Ledgers}
\end{figure*}

In order to collect information, it first has to listen to the entire channel it is interacting with in the current subframe. Upon receiving all packages and decoding, the Ledger information inside the package is added to the local Ledger to form the entire local Ledger, as Fig. \ref{fig:Ledgers} shows.

In order to collect whether a collision has occurred, the decoded Ledger information needs to be compared with the $L_A$ list obtained by sensing. The Ledger shows that the ID of the sub-channel occupied by the vehicle is the same as the corresponding sub-channel in the $L_A$ list, then the sub-channel of the sub-frame is normal. Conversely, if there is a sub-channel that shows busy in the $L_A$ list, but none of the packages show a vehicle occupying the sub-channel after decompression, then it is assumed that there is a collision in the corresponding sub-channel. In this case, we need to store the ID of the corresponding sub-channel and the information of the corresponding sub-frame in the newly formed Ledger. This will allow collision information to be shared when this package is broadcast.

\section{Ledger Structure}

For current BSM package, it is defined to include some safety notation, hazard alert, and other information. The Ledger presented in this paper contains not only this basic information, but also defines more information to help build a complete Ledger for other vehicles within the network.

In physical layer, it is significant to define multiple parameters. By the setup of each Ledger, the vehicle should package and transmit broadcast to the network with a certain payload size. The parameters in the physical layer should help to manage how many resources should be allocated to fully transmit one Ledger package. 

Each frame duration is 10 ms and each subframe duration is 1 ms, and the RB is defined as 12 consecutive subcarriers/Resource Elements(RE) in the frequency domain, which means $N_{sc}^{RB} = 12$ \cite{TS38211}. 

FR1 and FR2 are the two frequency bands established by 5G NR. The maximum bandwidth per carrier for FR1 is 100 MHz, with subcarriers of 15/30/60 kHz. For FR2, the maximum bandwidth per carrier is 400 MHz, with 60/120/240 kHz subcarriers. Clearly, as Table. \ref{tb:SCS} shows, a carrier's maximum bandwidth can affect the maximum number of packages that can be transmitted \cite{zhang2022multiaccess}.

As the physical layer defined, the time was divided into time slots, and the frequency domain was divided into subcarriers. In each slot, divided into 14 OFDM symbols. As Fig. \ref{fig:Ledgers} shows, one OFDM symbol with one subcarrier grouped one Resource Element (RE). 
\subsection{Timestamp}
In the 5G NR V2X, the protocol defines the frame structure in the physical layer to divide the time into frames and subframes. Each frame duration is 10 ms and each subframe duration is 1 ms. With different selection of numerology number $\mu$, the number of slot in each subframe duration was fixed. During each time slot, the vehicle could either receive packages or transmit one package due to the HD mode.

Each vehicle knows the current time by the Real Time Clock (RTC), and can calculated the timestamp when they will transmit this Ledger. There will have two situations:
\begin{itemize}
\item Within the SPS period
\begin{equation}\label{eq:timestamp1}
    t_{TimestampWithinSPS} = t_{last} + RRI
\end{equation}

$t_{TimestampWithinSPS}$ represent to the timestamp will be used for the new generated Ledger in this situation. $t_{last}$ means the last transmit slot start time.

\item Beginning of the SPS period
\begin{equation}\label{eq:timestamp2}
    t_{TimestampBeginSPS} = t_{current} + n_{select} * \tau_{slot, \mu}
\end{equation}

$t_{TimestampBeginSPS}$ represent to the timestamp will be used for the new generated Ledger in this situation. $ t_{current}$ means the current time from RTC. $n_{select}$ is the value decided by the selection window. In which time slot the vehicle will transmit the Ledger. The time duration of the slot is defined by the numorology $\mu$. Therefore, the timestamp of the package could be calculated as well.
\end{itemize}

This calculated time should included into the Ledger header, with corresponding position (timestamp) and length (10 bytes).

This timestamp is the most important part in the header to build up the entire Ledger.

\subsection{Vechicle ID}
In the V2V network, when the car entering this network, it will select a number to be its identification ID. There are two constrains for theis ID number: (1) never as same as other users in this network, and (2) length less than 1 byte. 

The selected ID value should be encapsulated into the corresponding position in the Ledger header. Once the ID is decided, it will never change before it leave the current network. 

\subsection{Sub-channel ID}
As the physical layer defined, one OFDM symbol and one subcarrier introduced one Resource Element (RE). Within one time slot, group $N_{sc}^{RB} = 12$ subcarriers derive one physical Resource Block (PRB). 


In NR V2X SL, $M_{sub}$ refers to how many PRBs contained in one sub-channel, which can be equal to 10, 12, 15, 20, 25, 50, 75, or 100. Each PRB could contain amount of bits, $N_{bits, MCS}$, which is depended on MCS. For SL,  PSSCH is assigned by Sidelink Control Information (SCI). The table for PDSCH could be used with MCS index $0 \leq I_{MCS} \leq 27$ \cite{TS38214}.

The UE shall first determine the number of REs ($N_{RE}^{PRB}$) within one PRB. Refer to the definition in \cite{TS38214}, the formula could be defined as:

\begin{equation}\label{eq:NRE'}
    N_{RE}^{PRB} = N_{sc}^{RB} ( N_{symbol}^{sh} - N_{symbol}^{PFSCH}) - N_{oh}^{PRB} - N_{RE}^{DMRS}
\end{equation}

As defined before, $N_{sc}^{RB}$ should be 12 for subcarrier in each RB. $N_{symbol}^{sh}$ with given formula $\mathit{sl\mbox{-}lengthSLsymbols}$ - 2, which return the value 12. $N_{symbol}^{PFSCH}$ is 0 because broadcast mode do not require any feedback information, therefore from the higher layer, parameter $\mathit{sl\mbox{-}PSFCH\mbox{-}Period}$ is 0. $N_{oh}^{PRB}$ and  $N_{RE}^{DMRS}$ both take zero. Refer to \cite{TS38214} definition, $N_{RE}^{DMRS}$ take 12. Therefore, the equation \ref{eq:NRE'} can be calculated to be 132. It means within one PRB, allocated 132 REs for PSSCH. 
 
According to table \ref{tb:SCS}, different $\mu$ decided the subcarrier spacing and the maximum carrier bandwidth ($CBW_{max}$). In frequency axis, it defined how many PRBs for one slot. It can defined as:

\begin{equation}\label{eq:PRBs}
    N_{PRB}^{slot} =  \floor*{ \dfrac{CBW_{max}}{SCS} * \dfrac{1}{12}}
\end{equation}

According to different MCS index, spectral efficiency will be different. For fixed payload of each package, defined one sub-channel in one slot could exactly enough for one package. If the original BSM payload ($P_b$) is 300 bytes, the total payload for the package with header information will be 350 bytes as the header size is 50 bytes. Here should define how many PRBs will need for one package transmission. 

\begin{equation}\label{eq:PRBs}
    N_{PRB}^{slot} = \floor*{ \ceil*{\dfrac{P_b *  8}{M_{order}} * \dfrac{1}{\eta}} * \dfrac{1}{N_{RE}^{PRB}}}
\end{equation}

$M_{order}$ is the modulation order which can be found in MCS table.

Obviously, the addition of the Ledger concept sacrifices some of the channel resources. Then, according to the design of this paper, it is possible to estimate the approximate extent of this resource loss. If the original BSM package contain 300 bytes information,  the new package with Ledger should be 350 bytes, which is 350 bytes =  350 * 8 = 2800 bits. By applying the fix modulation parameter: MCS choose 1, which is QPSK. By MCS table, it gives $\eta$ = 0.2344. Refer to equation \ref{eq:PRBs}, it is easy to have 2800/2/0.2344 = 5972.69624573, and choose ceiling to be 5973. After that, how many RPBs will be used to send a new package with Ledger can be calculated, which is 5973/132 = 46 PRBs. 

Then, by choosing the numerology $\mu$ = 0 for SCS = 15kHz, max carrier bandwidth will be 50MHz, and 1 slot means 1 ms. It can be found from Table \ref{tb:SCS}. 
Therefore, $N_{PRB}^{slot}$ = (50 * $10^6$) / (15 * $10^3$) / 12 = 277.77778, choose floor to be 277. As long as the PRBs per slot was calculated, it is easy to know how many sub-channels could have for every slot, $N_{sub-channel}^{slot}$ = 277/46 = 6.0217, choose floor to be 6. Thus, for RRI = 100 ms, can support maximum 600 cars. 

Now, for comparison, for original package without header, the package size will be 300 bytes and the value $N_{PRB}^{slot}$ = 7. It means, for RRI = 100ms, can support maximum 700 cars. According to the calculation, Ledger headers sacrifice 14.29\% of capacities. As the Fig. \ref{fig:scenario} shows, the maximum network will not contain up to 600 vehicles even if traffic congestion is present.  Therefore, it is not a big waste of resources for the V2V network.

\section{Simulation}

\begin{figure}
  \centering \includegraphics[scale=0.25]{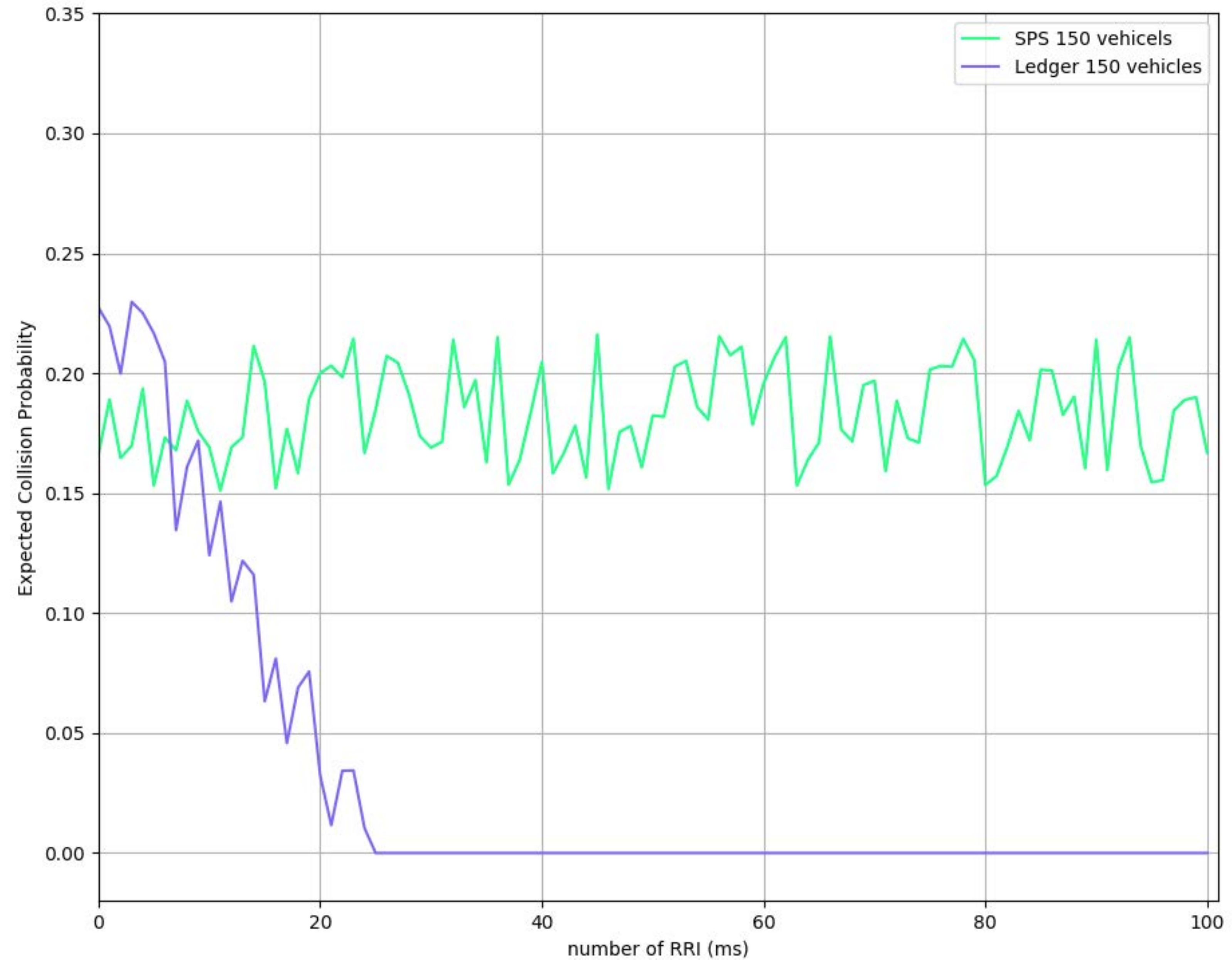}
  \caption{Ledgers vs. BSM.}
  \label{fig:simSPSvsLedger}
\end{figure}

\begin{figure}
  \centering \includegraphics[scale=0.25]{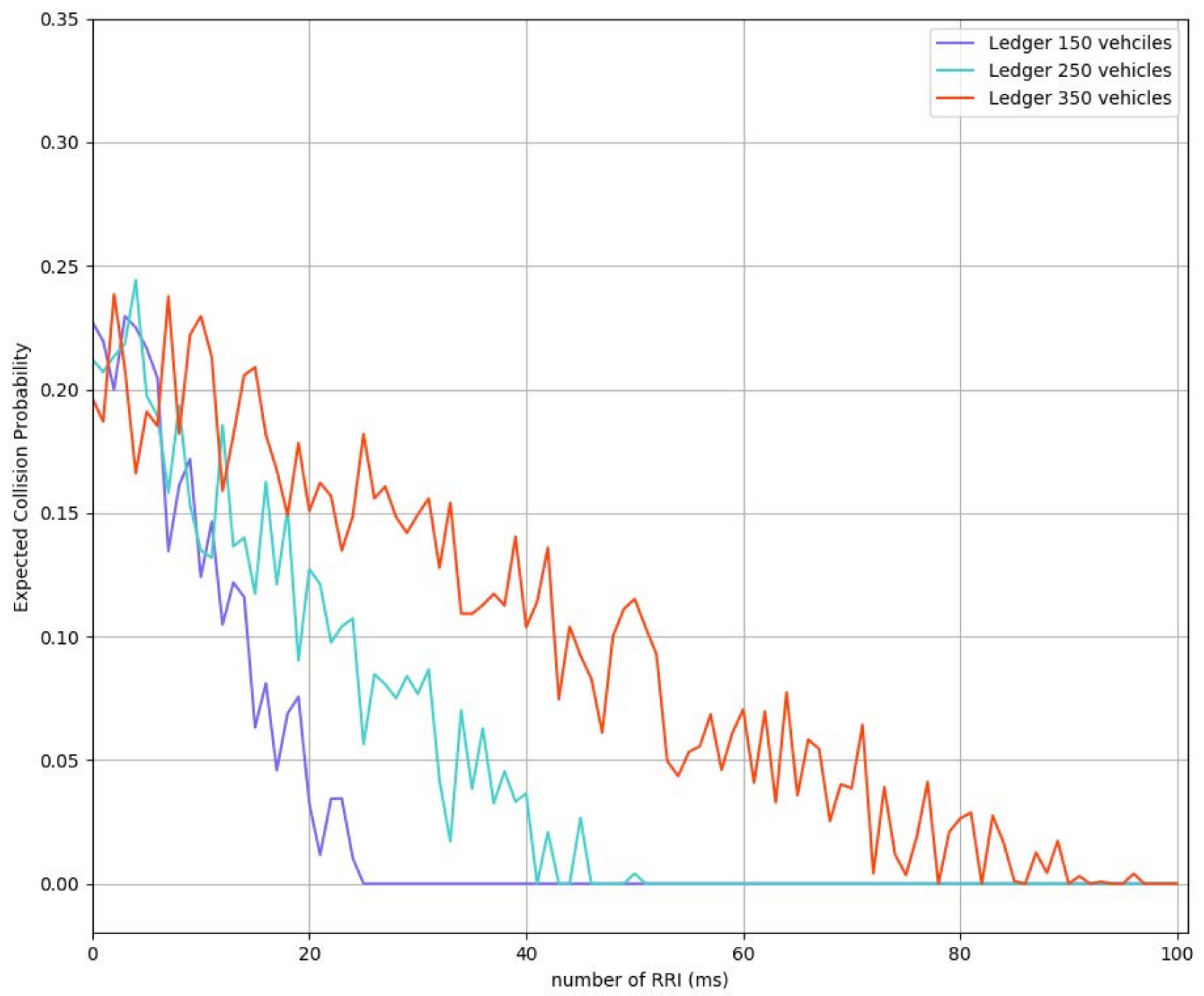}
  \caption{Ledgers based on different traffic load.}
  \label{fig:simLedgervsLedger}
\end{figure}

In the simulation, the Ledger was designed to be a linked structure. Each Ledger package contains the information designed as shown in \ref{fig:Ledgers}. This simulation compares the original BMS package and the Ledger design package. They both rely on the same traffic load and follow the SPS protocol. The simulation represent as Fig. \ref{fig:simSPSvsLedger}. Based on the Ledger design, the second simulation was designed to work with different traffic loads. The simulation result appears in Fig. \ref{fig:simLedgervsLedger}. 

The first simulation indicates that the collision probability of the original BSM package (same size 300 bytes) based on the SPS protocol is about the same. The reason is that the SPS follows random selection, and thus for each SPS period, there is a chance (1-$p$) to reselect a new sub-channel, even if there has not been a collision in the current sub-channel. In the Ledger design, the reselection probability will only appear when the collision happened. Mathematically, $p \in \{0, 1\}^n$. This means $p$ can only be 0 for keeping if there is no collision, and 1 for reselecting if a collision has occurred. $n$ is the value for number of SPS periods. 

The vehicle can detect if it collided with others by decoding the packages sharing in the network. Thus, by analyzing the simulation, it is easy to conclude that if there is a collision, the colliding vehicles could be aware of the collision within one RRI. This simulation shows that the colliding vehicle knows it is in collision between the time duration of [$T_{trans},  T_{trans}+100$]. $T_{trans}$ is the time at which the collision vehicle broadcast its package. 

The second simulation Fig. \ref{fig:simLedgervsLedger} was simulated based on the Ledger design. It showed the variation of performance based on different traffic load. One can observe that the higher traffic load, the longer time to converge. It is not obvious though, for the first 5 RRI, the performance of the Ledger design was the same as the original BSM packages. This is because the SPS protocol requirement. For each SPS period, the RC value control the time of each period. The range of RC is [5, 15]. As the result, the first 5 RRI, non of the vehicle could end the collision if the collision happened. After the first 5 RRIs, some of the vehicles start to reselect to have a new SPS period. Then based on the Ledger design, the keeping probability $p$ start to leading the reselection performance. The collide vehicle will start to reselect a new resource as soon as the end of the current period. Therefore, after the first 5 RRIs, the collision probability start to decrease fast.

The keeping probability $p$ also gives that if there is no collision, then the value will be 0. As a result, once the vehicle does not find collision, it will not change. In other words, other vehicles will never re-select this resource. Future collisions will be avoided and continuous collisions will be prevented. Thus, from the simulation results, after some RRIs, the collision probability converges to zero. The fixed resource will be used by every vehicle until it leaves the current network.

\section{conclusion}
With Ledger sharing, each vehicle can achieve full-duplex mode so that it can get collision information in one RRI time. At the end of the SPS cycle, the vehicle can have a clear decision to reselect a new resource or continue using the current one. Such a working mode allows each vehicle in a collision to select a new resource as soon as possible, thus avoiding the phenomenon of continuous collisions. Of course, if no collision arises from the initial selection, there will no longer be any possibility of collision without changing the resource.  After some RRIs, the collision probability can finally converge to 0.

For future works, this Ledger design could be improved to use RRI based reselection performance. The SPS protocol restricted vehicles to reselect after the end of the SPS period. However, the vehicle can be aware of whether it is colliding or not within one RRI. The rest of the RRIs in the SPS period will be a waste of time. If it is possible to reselect a distinct resource within one RRI if a collision occurs, the convergence time is possible to be much shorter.

\bibliographystyle{IEEEtran}
\bibliography{reference}

\end{document}